\def\aa{{A\&A}}
\def\aas{{A\&AS}}
\def\aj{{AJ}}
\def\annrev{{ARA\&A}}
\def\apj{{ApJ}}
\def\apjs{{ApJS}}
\def\baas{{BAAS}}
\def\mnras{{MNRAS}}
\def\nat{{Nature}}
\def\pasp{{PASP}}

\documentclass[cup5b]{caps}
\usepackage{graphicx}
\usepackage{amssymb}
\usepackage{ociwsymp4e}  %Use this for contributed papers.

\HeadText{Travaglio et al.}  %Enter author name; if three authors, list all three;
\begin{document}

\pagenumbering{arabic}

\author[]{C. TRAVAGLIO$^{1}$, K. KIFONIDIS$^{1}$, and E. M\"ULLER$^{1}$
\\
(1) Max-Planck Institut f\"ur Astrophysik, Karl-Schwarzschild Strasse 1,\\ 
D-85741 Garching bei M\"unchen, Germany}

\chapter{Multi-dimensional nucleosynthesis 
         calculations of Type~II SNe}

\begin{abstract}

We investigate explosive nuclear burning in core collapse supernovae
by coupling a tracer particle method to one and two-dimensional
Eulerian hydrodynamic calculations. Adopting the most recent
experimental and theoretical nuclear data, we compute the
nucleosynthetic yields for 15 $M_\odot$ stars with solar metallicity,
by post-processing the temperature and density history of advected
tracer particles. We compare our results to 1D calculations published
in the literature.

\end{abstract}

\section{Introduction}

The pre- and post-explosive nucleosynthesis of massive stars has been
studied extensively by several groups over the last years (see Woosley
\& Weaver~1995; Thielemann et al.~1996; Limongi et al.~2000; Rauscher
et al.~2002, and the references therein). Although a lot of work has
been performed in this field, computed nucleosynthetic yields are
still affected by numerous uncertainties. For instance, because of our
rather sketchy current understanding of the physical mechanism(s) that
lead from core collapse to supernovae (SNe), all studies of explosive
nucleosynthesis, that have been performed to date, made use of ad hoc
energy deposition schemes to trigger SN explosions in progenitor
models.  While the results of such calculations indicate that the
yields of only a rather small number of nuclei are sensitive to the
details of how the supernova shock is launched (see e.g. 
Woosley\&Weaver~1995), it is nevertheless important to attempt to compute 
nucleosynthetic yields in the framework of more sophisticated models of the 
explosion. The impact of multidimensional hydrodynamics has not been investigated 
in detail so far. In addition, among the isotopes whose yields are known to depend
sensitively on the explosion mechanism, and thus cannot be predicted
accurately at present, are key nuclei, like $\rm^{56}Ni$ and
$\rm^{44}Ti$, that are of crucial importance for the evolution of
supernova remnants and for the chemical evolution of galaxies. These
nuclei bare also important consequences for numerical supernova
models.  Their yields can be used as a sensitive probe for the
conditions prevailing in SNe and hence can serve to constrain
hydrodynamic SN models with their complex interdependence of
neutrino-matter interactions and multi-dimensional hydrodynamic
effects. This may ultimately aid in improving our understanding of the
explosion mechanism itself.

\section{Hydrodynamic models}

The nucleosynthesis calculations presented in this work are based on
one and two-dimensional hydrodynamic models of SNe which follow the
revival of the stalling shock, which forms after iron core collapse,
and its propagation through the star from 20 ms up to a few seconds
after core bounce (when the explosion energy has saturated and all
important nuclear reactions have frozen out). The simulations are
started from post-collapse models of Rampp \& Janka (priv. comm.), who
followed core-collapse and bounce in the 15 $M_\odot$, $Z = Z_\odot$
progenitors of Woosley \& Weaver~(1995) and Limongi et al.~(2000). We
employ the HERAKLES code, which solves the hydrodynamic equations in
1, 2 or 3 spatial dimensions with the direct Eulerian version of the
Piecewise Parabolic Method (Colella \& Woodward~1984), and which
incorporates the light-bulb neutrino treatment and the equation of
state of Janka \& M\"uller~(1996) (for more details see Kifonidis et
al.~2003, and the references therein). The main advantages of our
approach are that we drive the shock by accounting for neutrino-matter
interactions in the layers outside the newly born neutron star,
instead of using a piston (see e.g. Woosley \& Weaver~1995) or a
``thermal bomb'', and the possibility to perform calculations from one
up to three spatial dimensions. The main disadvantage is the fact that
we do not take into account the dense innermost layers of the neutron
star, and that we currently use a simplified scheme for neutrino
transport. Thus, we cannot obtain self-consistent neutrino
distribution functions, but have to assume the spectral
distribution of neutrinos and anti-neutrinos of all flavors that are
emitted by the neutron star, and the temporal evolution of their
luminosities. We assume the latter to be given by the simple
exponential law
\begin{equation}
  L_{\nu_i} = L^{0}_{\nu_i} {\rm e}^{-t/t_L},
             (\nu_i \equiv \nu_{\rm e}, \bar \nu_{\rm e}, 
               \nu_{\mu}, \bar \nu_{\mu},
               \nu_{\tau}, \bar \nu_{\tau})
\label{eq:Lumnu}
\end{equation}
where $t_L$ is of
the order of $700$\,ms, and the $L^{0}_{\nu_i}$ are parameters of the
calculation. The neutrino spectra are prescribed in the same way as in
Janka \& M\"uller~(1996).

\section{Marker particle method}

Choosing a hydrodynamic scheme for computing
multi-dimensional hydrodynamic models that include the
nucleosynthesis, one faces the dilemma of using either a Lagrangian
or an Eulerian method.  Since nuclear networks with hundreds of
isotopes are prohibitively expensive in terms of CPU time and memory
for multi-dimensional calculations, such networks can only be solved
in a post-processing step (the energy source term due to nuclear
burning can usually be calculated with a small network online with the
hydrodynamics, and may even be neglected completely in some cases,
depending on the structure of the progenitor). The Lagrangian approach
has the advantage that it naturally yields the necessary data for the
post-processing calculations, since it directly follows the evolution
of specific fluid elements. For the problem of neutrino-driven
supernovae, however, Lagrangian methods (like SPH) have a crucial
drawback: by their nature they concentrate resolution in
mass. Neutrino driven explosions, on the other hand, are triggered by
neutrino heating in a high-entropy, low-density region outside the
nascent neutron star. Failing to spatially resolve this region, which
contains only a small amount of mass, will lead to arguable
results regarding the hydrodynamics and thus ultimately to
doubtful nucleosynthetic yields. To achieve an adequate spatial
resolution, Eulerian schemes (where the grid is fixed in space) or
even adaptive schemes (in which the grid automatically adapts to
resolve steep gradients in the solution) are to be prefered. However, 
the problem then arises how one should obtain the necessary data for
the post-processing calculations. We do this by adding a ``Lagrangian
component'' to our Eulerian scheme in the form of marker particles
that we passively advect with the flow in the course of the Eulerian
calculation, recording their $T$ and $\rho$ history by interpolating
the corresponding quantities from the underlying Eulerian grid. A
similar method has been adopted in a previous study of
multi-dimensional nucleosynthesis in core collapse SNe by Nagataki et
al.~(1997), and more recently in calculations for very massive stars
(Maeda et al.~2002), and for Type~Ia SNe (Niemeyer et. al.~2002).

  \begin{figure}
    \centering
    \includegraphics[width=12cm,angle=0]{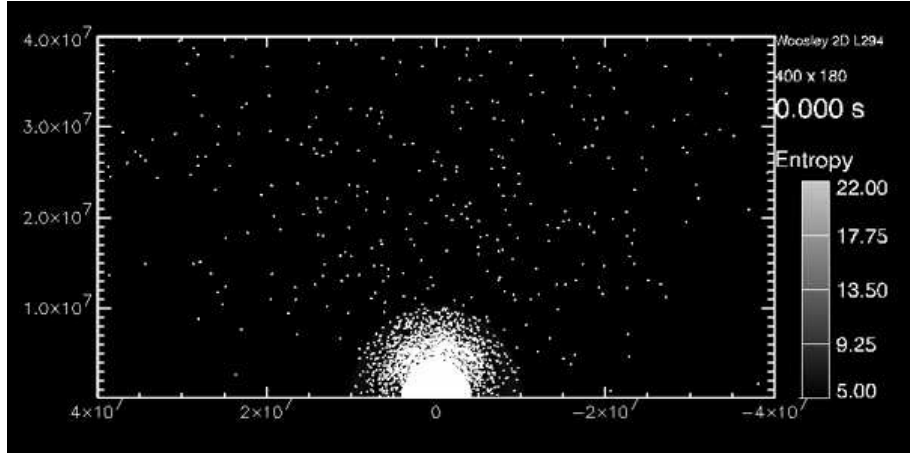}
    \caption{Initial marker particle distribution in the innermost 
             400\,km of the computational domain of a 2D simulation
             that was started from the Woosley \& Weaver~(1995) 
             progenitor. The entropy distribution 
             (in $k_b$/nucleon) is depicted in the background.}
    \label{fig:marker_inidist}
  \end{figure}

  \begin{figure}
    \centering
    \includegraphics[width=12cm,angle=0]{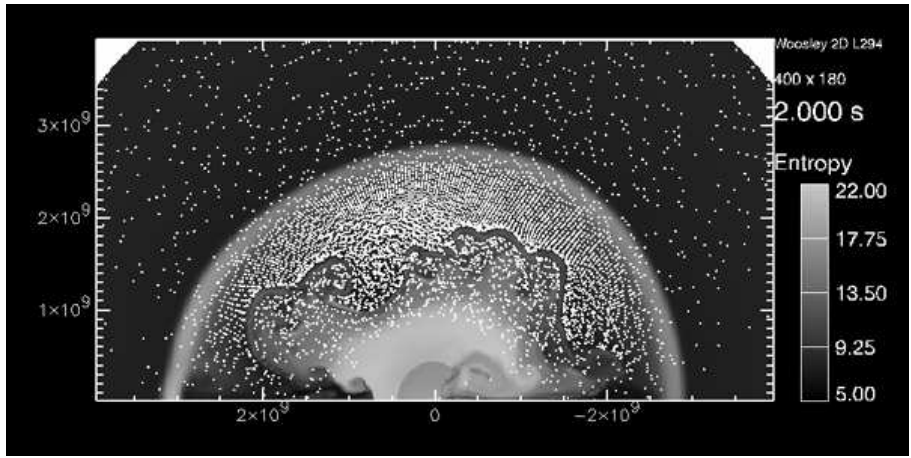}
    \caption{Final marker particle distribution of the 2D simulation 
             from \protect Fig.~\ref{fig:marker_inidist}. Note the change
             of the radial scale, and the pile-up of particles in the
             dense layer between the (light colored) low-density,
             neutrino-heated bubble, which is almost void of
             markers, and the shock farther out. The entropy distribution
             (in $k_b$/nucleon) is depicted in the background.}
    \label{fig:marker_findist}
  \end{figure}

For our 1D and 2D calculations we have used 1024 and 8000 marker particles, 
respectively. They are distributed homogeneously in mass throughout the
progenitor's Fe core, Si, O, and C shells assuming the composition of
the progenitor at the corresponding mass coordinate as the initial
composition of the respective tracer
particle. Figure~\ref{fig:marker_inidist} shows the initial
distribution of the particles in the innermost region of the
computational domain for a 2D simulation that was started from the
s15s7b progenitor of Woosley \& Weaver~(1995).  The final distribution
of the particles (at a time of 2\,s after core bounce) is given in 
Fig.~\ref{fig:marker_findist} for the same simulation. In both Figures 
the entropy distribution is plotted in the background. 
Figure~\ref{fig:marker_findist} demonstrates that the particles trace 
mainly the high-density region of the ejecta (which is located between the 
shock and the neutrino-heated bubbles), and that still the spatial 
resolution of the hydrodynamic calculation is not compromised in the 
low-density, neutrino-heated layers due to the Eulerian nature of our 
hydrodynamic scheme.

\section{Nucleosynthesis: first results and perspectives}

Given the temperature and density history of individual marker
particles we can calculate their nucleosynthetic evolution and compute
the total yields (including the decays of unstable isotopes) as a sum
over all particles. The reaction network employed for our
nucleosynthesis calculations contains 296 nuclear species, from
neutrons, protons, and $\alpha$-particles to $\rm^{78}Ge$
(F.-K. Thielemann, priv. comm.). The reaction rates include
experimental and theoretical nuclear data as well as weak interaction
rates. The $^{12}$C($\alpha$,$\gamma$)$^{16}$O rate is that of
Caughlan et al.~(1985). To investigate the sensitivity of the yields
with respect to the implementation of the nuclear physics, we have
also recalculated the nucleosynthesis with the nuclear network code of
M. Limongi (priv. comm.), and found good agreement with the results
obtained using the Thielemann network code.

  \begin{figure}
    \centering
    \includegraphics[width=9cm,angle=0]{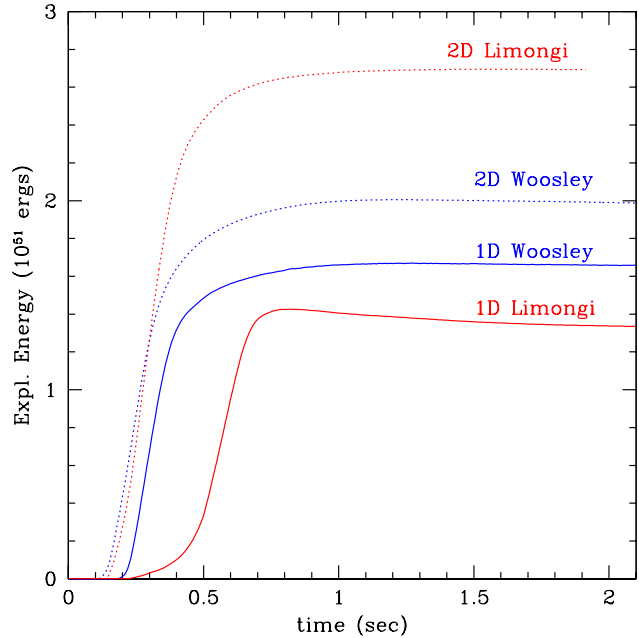}
    \caption{Explosion energies for diffferent models discussed
     in the text.}
    \label{fig:exp_energies}
  \end{figure}

In order to estimate the effects of the spatial resolution of the 
hydrodynamic calculations on the nucleosynthetic yields we have performed 
resolution studies in one spatial dimension. Varying both the number of 
markers and Eulerian zones, we adjusted the numerical resolution such that 
errors resulting from interpolation between these two ``grids'' are
less than a few per cent for a simulation with 2000 zones and 
1024 marker particles. Keeping the resolution of the Eulerian grid fixed 
at 2000 zones and varying the number of markers, we obtain convergence of
the yields, if the number of particles exceeds $\sim 1000$. For 10 times 
less markers, gradients in the hydrodynamic quantities are not sampled 
sufficiently accurately, affecting the final composition by $\sim$20\%.  
Numerical convergence depends also on the accuracy of the
hydrodynamic quantities themselves, i.e. on the resolution of the
Eulerian grid. We have not investigated this in detail so far but
plan to do this in forthcoming calculations. In addition, the
one-dimensional results may not be applicable to the two-dimensional
situation. Therefore, a resolution study in two spatial dimensions is 
also in preparation.

So far we have investigated four explosion models for their
nucleosynthetic yields: a one-dimensional and a corresponding
two-dimensional model that made use of model s15s7b of Woosley \&
Weaver~(1995) and a second pair of a one and two-dimensional simulation
for the 15 $M_\odot$ Limongi et al.~(2000) progenitor.  The properties
of these models are given in Table~1, where $L^0_{\nu_e,52}$ is the electron 
neutrino luminosity (in units of 10$^{52}$ erg/s), $E_{exp,51}$ is the explosion
energy (in units of 10$^{51}$ erg), and $t_{exp}$ is the explosion time scale
(in ms) defined as the time after the start of the simulation when the explosion
energy exceeds 10$^{49}$ erg (for a detailed explanation of the neutrino 
parameters see Janka \& M\"uller~1996 and Kifonidis et al.~2003).  

 \begin{table}
  \caption{Parameters of models, using the Woosley\&Weaver~(1995, WW95) and 
           Limongi et al.~(2000, LSC00) progenitors.}
    \begin{tabular}{c|ccccc}   
     \hline \hline
     Model & Zones & $N_{markers}$ & $L^0_{\nu_e,52}$ & $E_{exp,51}$ & $t_{exp}$ (ms) \\
     \hline
     1D WW95  & 2000           & 1024 & 2.940 & 1.46 & 230 \\
     2D WW95  & 400$\times$180 & 8000 & 2.940 & 1.99 & 125 \\
     1D LSC00 & 2000           & 1024 & 3.365 & 1.33 & 260 \\
     2D LSC00 & 400$\times$180 & 8000 & 3.365 & 2.69 & 150 \\
     \hline \hline
    \end{tabular} 
  \label{sample-table}
\end{table}

Figure~\ref{fig:exp_energies} shows the evolution of the explosion energy for 
the four models, using the same neutrino luminosity for the 1D and 2D
model of the same progenitor. For both progenitors the 2D model explodes 
with higher energy than the corresponding 1D one. The Limongi et al.~(2000) 
progenitor needs higher neutrino luminosity to explode, mainly due to the fact 
that it has a more compact core.

  \begin{figure}
    \centering
    \includegraphics[width=9cm,angle=-90]{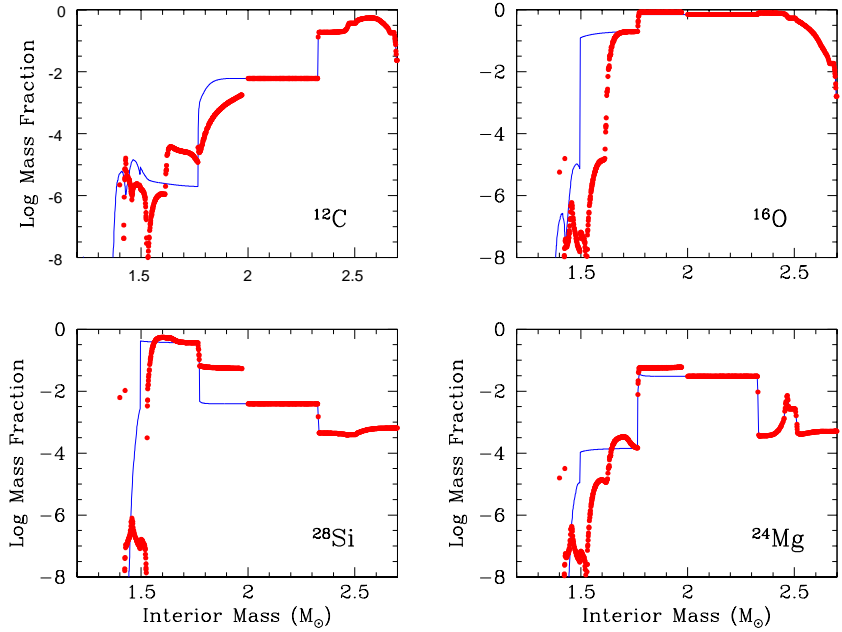}   
    \caption{Logarithm of the final mass fractions ({\it filled dots}) of 
             $^{12}$C,$^{16}$O,$^{24}$Mg, and $^{28}$Si as a function of mass
             coordinate for our 1D nucleosynthesis calculation
             employing the  15 $M_\odot$ Woosley \& Weaver~(1995)
             progenitor. Each dot represents a marker particle. The
             presupernova composition is also plotted ({\it thin line}).}
    \label{fig:mass_frac_1D}
  \end{figure}

In Fig.~\ref{fig:mass_frac_1D} we show the final mass fractions of 
$^{12}$C, $^{16}$O, $^{24}$Mg, and $^{28}$Si for the 1D calculation that 
made use of the Woosley \& Weaver~(1995) progenitor. Each dot represents a 
marker particle that is located at a certain mass coordinate and carries a 
specific composition. As a reference, we also plot the presupernova composition
(Woosley \& Weaver~1995). Comparing our 1D yields with the explosive 
nucleosynthesis results of Woosley \& Weaver~(1995), we find
very good agreement (with small deviations of the order of few
per cent) for the light elements. The position of the mass cut at 1.28 $M_\odot$ 
(which is obtained by comparing the velocity of each tracer particle 
with the local escape velocity) is in good agreement with the Woosley \& Weaver~(1995)
result, too.

\begin{figure}
    \centering
    \includegraphics[width=11cm,angle=0]{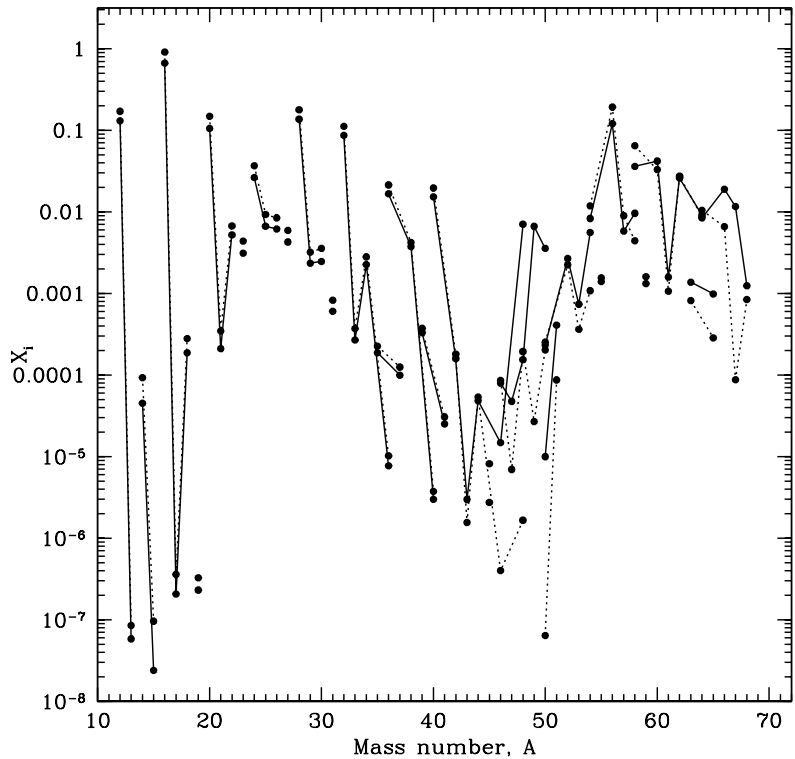}
    \caption{Final mass fractions obtained for the 1D ({\it dotted
             line}) and 2D simulation ({\it solid line}) as a
             function of the atomic number, using the 15 $M_\odot$
             Woosley \& Weaver~(1995) progenitor.}
    \label{fig:yields_1D-2D}
  \end{figure}

In Fig.~\ref{fig:yields_1D-2D} we compare the yields of the 1D and 2D
simulation for the Woosley \& Weaver~(1995) progenitor. The differences, 
which are apparently negligible in case of the lighter nuclei and small for 
the heavier ones, are mainly due to the on average higher temperatures in the 
2D simulation, i.e. more free neutrons are available in the innermost layers 
of the 2D simulation. This results in higher production factors for isotopes 
which are very sensitive to neutron captures, like e.g. $^{46,48}$Ca, 
$^{49,50}$Ti, $^{50,51}$V, $^{54}$Cr, and $^{67}$Zn.

The reason for the rather small differences in the yields between the 1D
and 2D simulation are the high initial neutrino luminosities, that we
adopted for our calculations, and their rapid exponential decline.
This leads to very rapid (and energetic) explosions 
(Fig.~\ref{fig:exp_energies}).  The short explosion time scale prevents
the convective bubbles, which form due to the negative entropy
gradient in the neutrino-heated region, to merge to large-scale
structures that can lead to global anisotropies, and hence to
significant differences compared to the 1D case.  Lowering the
neutrino luminosities (and the explosion energies), we obtain stronger
convection that strongly distorts the shock wave by developing large
bubbles of neutrino-heated material (see Janka \& M\"uller~1996;
Kifonidis et al.~2000; Kifonidis et al. 2001; and Janka et
al.~2001 for examples).  Adopting constant core luminosities instead 
of the exponential law of Eq.~(\ref{eq:Lumnu}), we can produce models 
where the phase of convective overturn lasts for several turn-over times 
and which exhibit the vigorous boiling behaviour reported by Burrows et 
al.~(1995). Such cases can finally develop global anisotropies, showing a 
dominance of the $m=0$, $l=1$ mode of convection (see Janka et al.~2003; 
Scheck et al., in preparation). As a consequence, convection can lead to 
large deviations from spherical symmetry, and thus to larger differences 
in the final yields than those visible in Fig.~\ref{fig:yields_1D-2D}.  
We are currently investigating such models in more detail.

\section{Electron captures in explosive conditions}

A common result of nucleosynthesis studies that have been performed to
date is that weak-interactions (in particular electron captures) do
not play an important role for explosive nucleosynthesis conditions
(i.e. for temperatures and densities obtained from
artificially induced 1D explosions of Type~II SN progenitors). In
contrast, electron captures are known to be very important in
presupernova hydrostatic nucleosynthesis (see e.g. the discussion in 
Woosley \& Weaver~1995).

  \begin{figure} \centering
    \includegraphics[width=9cm,angle=-90]{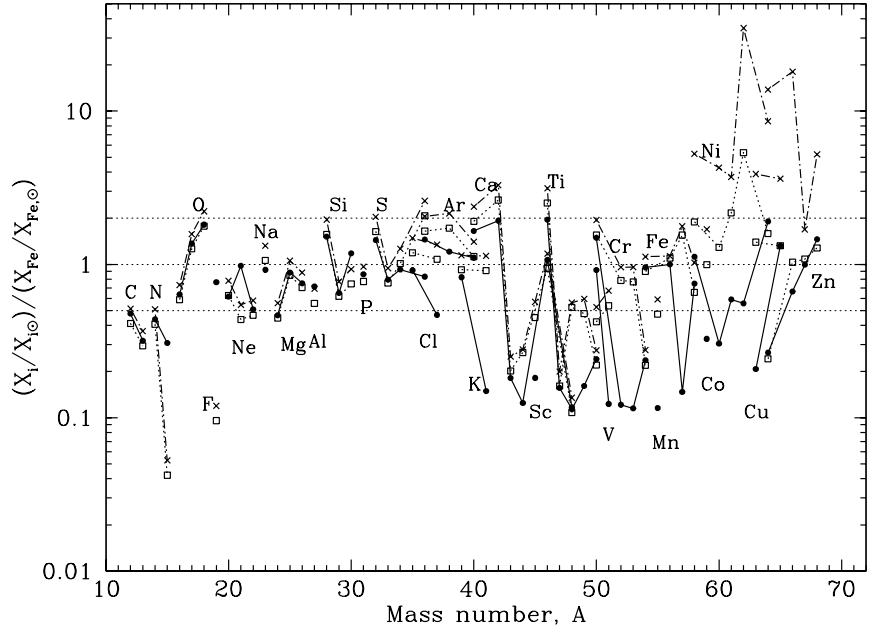}
    \caption{Mass fractions normalized to solar and
    to Fe abundance obtained for our 1D simulation using the Woosley
    \& Weaver~(1995) progenitor. The Woosley \& Weaver yields are
    given by {\it solid lines} and {\it filled dots}. Our 1D 
    simulation results including weak-interactions are shown by {\it
    dot-dashed lines} and {\it crosses}.  Our 1D simulation results 
    excluding weak-interactions are given by {\it dotted lines} and 
    {\it open squares}.}  \label{fig:e-captures} \end{figure}

\begin{figure}
    \centering
    \includegraphics[width=10cm,angle=0]{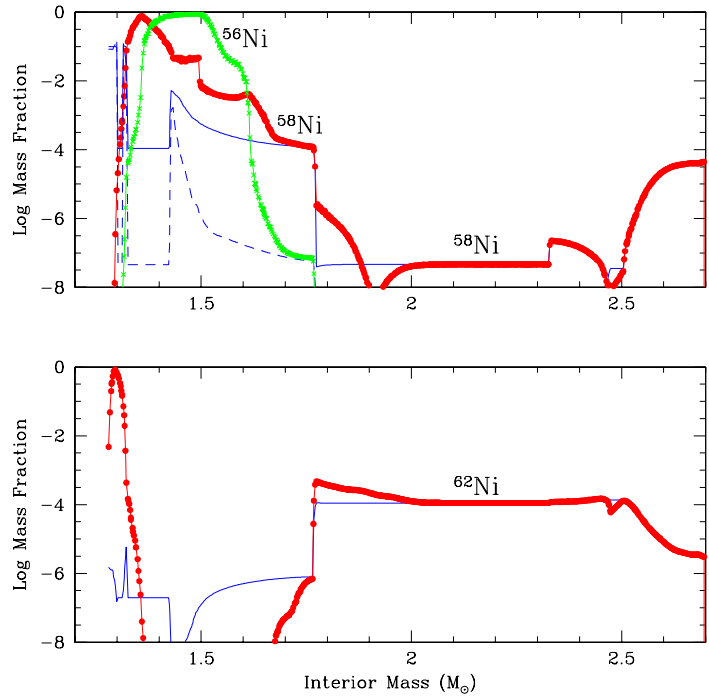}
    \caption{Logarithm of the mass fractions of   
             $^{56}$Ni and $^{58}$Ni ({\it upper panel}) and $^{62}$Ni 
             ({\it lower panel}) as a function of mass
             coordinate for our 1D nucleosynthesis calculation 
             employing the  15 $M_\odot$ Woosley \& Weaver~(1995)     
             progenitor. Each {\it dot} (for $^{58}$Ni and $^{62}$Ni)
             and {\it cross} (for $^{56}$Ni) represents a marker particle. 
             The presupernova composition is also plotted ({\it thin solid and 
             dashed line}).}
    \label{fig:massfrNi}
  \end{figure}

In Fig.~\ref{fig:e-captures} we summarize the effect of electron
captures on the yields of our 1D simulation that employs the 
Woosley \& Weaver progenitor.  The fact that markers with
temperatures of $\sim 8\times10^9$ K encounter high densities 
($\geq 10^8$ g/cm$^3$) leads to a large overproduction of 
neutron-rich Ni isotopes like $^{58}$Ni and $^{62}$Ni (from 100 up
to 1000 times the solar value).  According to our simulation 
(Fig.~\ref{fig:massfrNi}) markers with this composition are located at 
a mass coordinate $\sim$1.3 $M_\odot$ and have a typical Y$_e$ of 
$\sim$0.48. As their velocity is higher 
than the local escape velocity, they have to be considered in the calculation 
of the final yields (unless there occurs late fallback, see below).

Similar effects regarding the production of neutron-rich Ni isotopes
as in the 1D case are found in the 2D model (see also the
discussion regarding the differences between the 1D and 2D models in
the previous Section). In addition, the 2D model causes a very high
production of neutron-rich isotopes like $^{46,48}$Ca, $^{49,50}$Ti,
$^{50,51}$V, $^{54}$Cr, and $^{67}$Zn. Markers enriched in these 
isotopes are located in the innermost ejecta and have the lowest
Y$_e$ ($\sim$0.45). The fact that we obtain a high production of
these isotopes only in the 2D model and not in 1D it is due to
the higher temperatures reached in the innermost layers of the 2D 
simulation. This causes a higher degree of neutronization of these
regions.

As we already mentioned it is possible that the overproduction
problem might be resolved, at least in part, by late fallback.  This,
however, will most likely be only a viable solution for models with
explosion energies that are significantly smaller than the ones that
we have discussed here.  Whether or not the overproduction problem can
be solved will in addition depend on the efficiency of Rayleigh-Taylor
mixing during the late-time evolution.  Several episodes of
deceleration (accompanied by the formation of reverse shocks and by
Rayleigh-Taylor instabilities) are known to occur in the ejecta of
Type~II SNe when the supernova shock slows down in the He core and in
the H envelope of the progenitor (see Kifonidis et al.~2003 for
details). If most of the neutron-rich isotopes are located in the
high-entropy, low-density neutrino-heated bubbles they will indeed
have a higher probability to fall back to the core later on, because
they will not be able to participate very efficiently in the
Rayleigh-Taylor mixing at the Si/O and O/He interfaces farther
out. This mixing leads to the formation of clumps that decouple from
the flow and move ballistically through the ejecta (Kifonidis et
al.~2003) and thus make the material in the clumps less prone to
fallback. More detailled conclusions, however, can only be drawn with
a larger number of models that have to follow also the late-time
evolution of the ejecta.

\section{Conclusions}

We have presented a {\it marker particle} method to calculate multi-dimensional 
explosive nuclear burning in core collapse supernovae using a nuclear network of 
$\sim$300 isotopes. We have discussed one- and two-dimensional hydrodynamic
models of SNII that were computed starting from 15 $M_\odot$ progenitors with 
solar metallicity (Woosley \& Weaver~1995; Limongi et al.~2000).
With the temperature and density hystory of individual tracer particles, we 
presented and discussed the nucleosynthesis we obtained for these models,
comparing 1D and 2D calculations. In particular we pointed out the sensitivity
of the results to the neutrino luminosities (i.e. explosion energy) used in the 
hydrodynamic simulations. Different models with different explosion energies
are currently under investigation.
Finally we pointed out the need of late-time calculations of the evolution of 
the ejecta in order to better evaluate the amount of fallback and therefore
the yields.

%Instead of \verb"\includegraphics", figures can be inserted with 
%\verb"\plotone{fig1.ps}" or \verb"\plottwo{fig1a.ps}{fig1b.ps}".

\def\aa{{A\&A}}
\def\aas{{A\&AS}}
\def\aj{{AJ}}
\def\annrev{{ARA\&A}}
\def\apj{{ApJ}}
\def\apjs{{ApJS}}
\def\baas{{BAAS}}
\def\mnras{{MNRAS}}
\def\nat{{Nature}}
\def\pasp{{PASP}}

\end{document}